\documentclass[epjCONF, onecolumn]{svjour}
\usepackage{graphics}
\usepackage[varg]{txfonts} 
\usepackage[latin1]{inputenc}
\session-title{Assembling the puzzle of the Milky Way}
\begin{document}
\title{Two views of globular cluster stars in the Galactic halo}
\author{Sarah L. Martell\inst{1}}
\institute{Astronomisches Rechen-Institut\\
Zentrum f\"{u}r Astronomie der Universit\"{a}t Heidelberg\\
M\"{o}nchhofstrasse 12-14\\
69120 Heidelberg, Germany\\
\email{martell@ari.uni-heidelberg.de}
}
\abstract{
In Martell \& Grebel (2010)\cite{MG10} we reported the discovery in the Sloan Digital Sky Survey-II/SEGUE spectroscopic database of a small subset of halo red giants, $2.5\%$, with CN and CH band strengths indicative of globular-cluster-like carbon and nitrogen abundances. Because the formation of stars with unusual light-element abundances is thought to be restricted to high-density environments like globular clusters, this result has strong implications for both cluster formation processes and the assembly history of the Galactic halo. Here we discuss two efforts to expand upon that work.
} 
\maketitle
\section{Introduction}
\label{intro}
Globular clusters in the Milky Way all contain clear light-element
abundance variations (e.g., Carretta et al. 2009\cite{C09}; Pancino et
al. 2010\cite{P10}; Smolinski et al. 2011\cite{SMB11}, Martell 2011\cite{M11}). The typical pattern is C-N, O-Na
and Mg-Al anticorrelations, with roughly half of the stars having
scaled-Solar abundances and half showing a range of depletions in C, O
and Mg and enhancements in N, Na and Al. This pattern is presently
interpreted as a sign that stars in globular clusters are formed in
two closely spaced generations. The variations in light-element
abundance pattern (and the lack of $\alpha$-element or iron
enhancement between the two generations) are then thought to be a
result of chemical feedback only in the light elements, through the
retention and recycling of winds from AGB stars (e.g., D'Ercole et
al. 2008\cite{DVD08}), rapidly rotating massive stars (e.g., Decressin
et al. 2010\cite{D10}), lossy mass transfer in massive binary systems
(e.g., de Mink et al. 2009\cite{dM09}) or some combination of these sources. 

Although light-element abundance variations are found in all Galactic
globular clusters, they have not been found in open clusters (e.g.,
Martell \& Smith 2009\cite{MS09}; de Silva et al. 2009\cite{dS09};
Jacobson et al. 2008\cite{JF08}),
lending weight to the idea that the high density of early globular
clusters permitted a partial self-enrichment, and indicating that
unusual light-element abundances can be used to identify stars that
originally formed within globular clusters, even after those clusters
have been disrupted through tidal interactions with the Galaxy or
evaporated through internal two-body interactions. 

\section{Cluster-originating stars in SDSS data}
\label{sec1}
Martell et al. 2011\cite{MSB11} extended the search for stars
with globular cluster-like light-element abundances to the SEGUE-2
data set (Eisenstein et al. 2011\cite{SDSS3}), a low-resolution
spectroscopic survey based on SDSS photometry that targeted red giant
stars to larger distances than the original SEGUE survey (Yanny et
al. 2009\cite{SEGUE}). The authors' main goal was to search for trends
between Galactocentric distance and the frequency of stars originating
in globular clusters. They confirmed the overall result of Martell \&
Grebel (2010)\cite{MG10}, finding that 16 of the 561 halo red giant in their
sample, roughly $3\%$, were CN-strong and CH-weak relative to the
typical star at fixed metallicity and evolutionary phase, indicating
depleted carbon and enhanced nitrogen abundance. They also found that
the CN-strong stars are slightly more centrally concentrated in their
sample than the CN-normal stars, with very few CN-strong stars found
beyond a Galactocentric distance of 30 kpc. There are several recent
theoretical models of galaxy formation (e.g., Oser et
al. 2010\cite{O10}; Font et al. 2011\cite{F11}) that predict two-population halos in spiral galaxies,
with the inner halo formed mainly \textit{in situ} and the outer halo
formed primarily through accretion of satellites. If this result can be
confirmed in future surveys of Galactic chemodynamics, it will
implicate globular clusters as an important site of star formation in
the early Galaxy.

\section{Determining the full light-element abundance pattern}
\label{sec2}
\begin{figure}
\resizebox{0.75\columnwidth}{!}
{\includegraphics{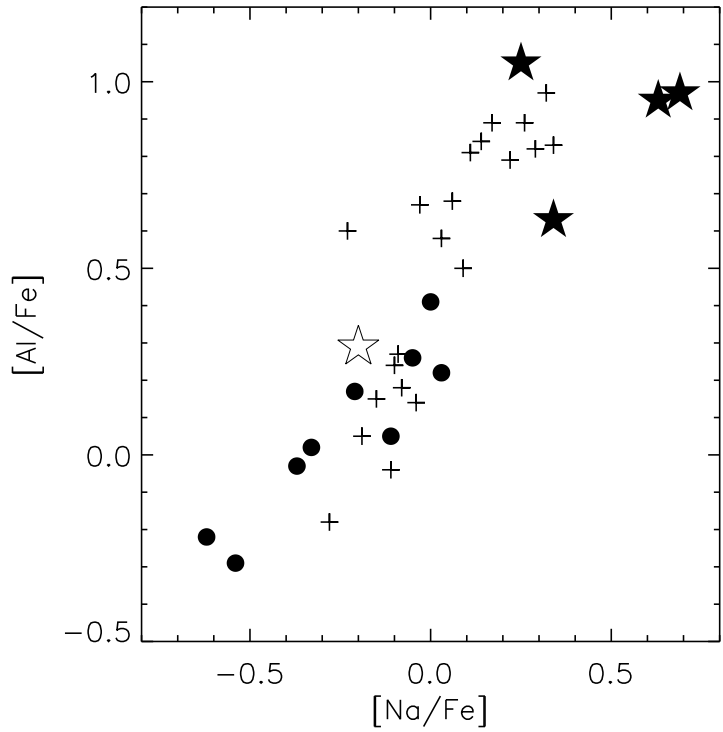} }
\caption{Sodium and aluminium abundances for moderate-metallicity
  CN-strong (filled stars) and CN-normal (open star) stars from our
  high-resolution followup observations at HET, along with
  similar-metallicity globular cluster (crosses) and field stars
  (filled circles) from the literature.}
\label{ff1}       
\end{figure}

High-resolution spectroscopic followup observations of the CN-strong
SEGUE stars has been underway since early 2010, using the
High-Resolution Spectrograph (Tull 1998\cite{HRS-HET}) on the Hobby-Eberly Telescope at
McDonald Observatory and the HIRES spectrograph (Vogt et al. 1994\cite{HIRES}) at Keck
Observatory (Martell, Shetrone \& Lai 2011, in prep). A total of 29
CN-strong stars selected from Martell \& Grebel (2010)\cite{MG10} have been
observed at resolutions between $\Delta \lambda/\lambda$ of $15,000$
and $30,000$, along with 16 CN-normal stars, also from Martell \&
Grebel (2010)\cite{MG10}, which were observed as a control set. Analysis of the
spectra is ongoing, and the preliminary results are promising. Figure
\ref{ff1} shows sodium and aluminium abundances for four
moderate-metallicity CN-strong stars (filled stars) and one CN-normal
star (open star) from our followup data obtained at HET, along with
comparison stars of similar metallicity: red giants in the globular
cluster M3 (crosses, data from Sneden et al. 2004\cite{S04}) and field giants
with metallicity similar to M3 (filled circles, data from Fulbright 2000\cite{F00}). While the M3 stars cover a wide, correlated range in sodium and
aluminium abundance, the field stars are restricted to lower [Na/Fe]
and [Al/Fe] abundances. The CN-normal field star from our new
observations falls together with the other field stars at the
low-abundance end of the sequence, and the CN-strong stars all fall at
the high-abundance end of the sequence, indicating that the CN-CH
selection employed in Martell \& Grebel (2010)\cite{MG10} and Martell
et al. (2011)\cite{MSB11} is effective for identifying stars with the full anomalous
light-element abundance pattern from C through Al.

\end{document}